\documentclass[11pt,a4paper]{article}
\usepackage{graphics}
\usepackage{epsfig}
\usepackage{hhline}
\usepackage{rotating}

\title{The Conformal Group $SO(4,2)$ 
and Robertson-Walker spacetimes}
\author{}
\author{Aidan J. Keane and Richard K. Barrett\footnote{Department of Physics 
and Astronomy, University of Glasgow, Glasgow G12 8QQ, Scotland, 
UK. tel: +44 141 339 8855 x0147}}

\newcommand{\be}{\begin{equation}}
\newcommand{\ee}{\end{equation}}
\newcommand{\ba}{\begin{eqnarray}}
\newcommand{\ea}{\end{eqnarray}}
\newcounter{saveeqn}

\begin{document}
\maketitle

\abstract{The Robertson-Walker spacetimes are conformally flat and so are 
conformally invariant under the action of the Lie group $SO(4,2)$, the 
conformal group of Minkowski spacetime. We find a local coordinate 
transformation allowing the Robertson-Walker metric to be written in a  
manifestly conformally flat form for all values of the curvature parameter $k$ 
continuously and use this to obtain the conformal Killing vectors of the 
Robertson-Walker spacetimes directly from those of the Minkowski spacetime.
The map between the Minkowski and Robertson-Walker spacetimes preserves 
the structure of the Lie algebra $so(4,2)$.
Thus the conformal Killing vector basis obtained does not depend upon $k$, 
but has the disadvantage that it does not contain explicitly a basis for the 
Killing vector subalgebra.
We present an alternative set of bases that depend (continuously) on $k$ 
and contain the Killing vector basis as a sub-basis (these are compared 
with a previously published basis).
In particular, bases are presented which include the Killing vectors 
for all Robertson-Walker spacetimes with additional symmetry, including the 
Einstein static spacetimes and the de Sitter family of spacetimes, where the 
basis depends on the Ricci scalar $R$.}

\vskip\baselineskip

\section{Introduction}
We are interested in the geometrical symmetry properties of the 
Robertson-Walker spacetimes. These models are a special case of the general 
conformally flat spacetimes, or to be precise, conformally Minkowski spacetimes.
The conformal group of Minkowski spacetime is the Lie group $SO(4,2)$. Thus 
any spacetime conformally related to Minkowski spacetime will be conformally 
invariant under the action of $SO(4,2)$.  
As a result, every conformally flat spacetime will admit a Lie algebra of 
conformal Killing vectors, that Lie algebra being $so(4,2)$. 
The Robertson-Walker spacetimes are spatially homogeneous and isotropic and 
so admit, at least, a 6-dimensional Lie algebra of Killing vector fields.
In some cases there is additional symmetry i.e. the Einstein static spacetimes 
and the de Sitter family of spacetimes.

Levine [1] discusses isometries and conformal motions in conformally flat 
spaces and determines the conditions which such a space must satisfy in order 
that it may admit an isometry group. Levine [2] classifies conformally flat 
spaces according to these conditions.
Maartens and Maharaj [3] have obtained a basis for the conformal Killing vector algebra of Robertson-Walker space-times for $k \pm 1$  by 
generalising a Minkowski conformal Killing vector and then commuting this 
with the Killing vectors to generate a complete set. 
However, some of the commutation relations presented in the paper are 
incorrect for the cases $k= \pm 1$ and we have rectified this. 

In this paper we derive the conformal Killing vectors of the Robertson-Walker 
spacetimes: we find a coordinate system in which the general Robertson-Walker 
metric takes on a manifestly conformally flat form in which case the conformal 
Killing vectors of the Robertson-Walker spacetimes are simply the conformal 
Killing vectors of the Minkowski spacetime.
The transformation we presnet is valid for all values of curvature $k$ 
continuously. 
We discuss the properties of the conformal Killing vectors and the structure of 
the Lie algebra $so(4,2)$ and present alternative bases according to 
the value of $k$.
We show why the basis presented in Maartens and Maharaj [3] in the case 
$k= \pm 1$ is only valid for these cases. 
We then present a basis which contains a basis for the Killing vector 
subalgebra for general Robertson-Walker models and which is legitimate for 
{\it all} values of curvature $k$ continuously. 
The de Sitter family of spacetimes, which are completely homogeneous and 
isotropic in four dimensions, are a special case of the general Robertson-Walker spacetimes.
These can similarly be written in a conformally flat form characterised by 
the (constant) Ricci scalar $R$.
Thus we can make similar statements regarding the conformal structure of 
these spacetimes.

There are a number of reasons why it is advantageous to have a basis for the
algebra that explicitly contains a sub-basis for the Killing vectors.
Firstly, it enables the dimension of the isometry group to be
gleaned immediately in any case by examining the conformal factors
(Tables \ref{tab:frwckv} and \ref{tab:frwckv-newbasis}). Perhaps more
important, though, is the fact that in any application where symmetry
information is useful, such as the study of particle motion
(as in Maharaj and Maartens [4] and Maartens and Maharaj [5]) or of 
perturbations of Robertson-Walker spacetimes, the ability to easily identify 
the Killing vectors (which give rise to conserved quantities) can greatly 
simplify the problem.

In the following greek indices take on the values $0,1,2,3$ and latin 
indices $1,2,3$, unless otherwise indicated.   

\section{Conformal Killing vectors}
Let $M$ be a spacetime manifold with metric tensor ${\bf g}$ of Lorentz 
signature. Any vector field ${\bf X}$ which satisfies 
\be
{\cal L}_{\bf X} \, {\bf g}=2 \phi(x^\alpha) \, {\bf g} \label{eqn:ckv}
\ee  
is said to be a {\it conformal Killing vector} (CKV) of ${\bf g}$. 
If $\phi$ is not constant on $M$ then ${\bf X}$ is called a {\it proper 
conformal Killing vector}, if $\phi$ is constant on $M$ then ${\bf X}$ is 
called a {\it homothetic Killing vector} (HKV) and if $\phi$ is constant and $\phi \ne 0$ on $M$ then ${\bf X}$ is called {\it proper homothetic}.
If ${\phi}_{;\alpha\beta} = 0$ then ${\bf X}$ is called a {\it special 
conformal Killing vector} (SCKV) and if $\phi=0$ then ${\bf X}$ is said to be a 
Killing vector (KV). 

The maximum dimension of the algebra of CKV on $M$ is 15 and this is 
achieved if $M$ is conformally flat. The maximum dimension of the KV algebra is 
10 and this occurs if $M$ is of constant curvature. If $M$ is not of constant 
curvature then this algebra has dimension at most 7. The algebra of HKV has 
dimension equal to or at most one greater than that of the KV algebra. 
If this algebra has its maximum dimension of 11 then $M$ is flat. 
Any CKV field in a flat spacetime is a SCKV and so the maximum dimension of 
the SCKV algebra is 15. If this occurs $M$ is flat whilst if $M$ is non-flat 
its maximum dimension is 8. For details and proofs see Hall [6] and 
references therein. It will be assumed throughout this paper that the 
spacetimes considered admit no local (nonglobalizable) conformal Killing vector 
fields.

Now, if we have a conformally related metric tensor $f^2(x^\alpha){\bf g}$ then 
we can investigate the Lie derivative of this metric tensor with respect to 
the CKV of ${\bf g}$, defined in (\ref{eqn:ckv}). We have that
\ba
{\cal L}_{\bf X}(f^2{\bf g}) & = &
{\cal L}_{\bf X}(f^2){\bf g}+f^2{\cal L}_{\bf X}({\bf g}) \nonumber\\
& = & {\bf X}(f^2){\bf g}+2 f^2 \phi \, {\bf g} \nonumber\\ 
& = & 2 \, [{\bf X}(f^2) / 2 f^2 + \phi] \, (f^2 {\bf g}) \nonumber\\
& = & 2 \, \phi'(x^\alpha) \, (f^2 {\bf g}) \: \: .
\ea
It is now obvious that a CKV ${\bf X}$ of ${\bf g}$ is necessarily a CKV of $f^2{\bf g}$ with conformal factor \newline 
$\phi'=[{\bf X}(f^2) / 2 f^2 + \phi]$.

Let us investigate the variation of the quantities ${\bf X} \cdot {\bf p}$ 
along a geodesic with tangent vector ${\bf p}$. It is straightforward to show 
that 
\be 
\nabla_{\bf p} ({\bf X} \cdot {\bf p}) =  \phi \; 
{\bf g}({\bf p},{\bf p}) \: \: .
\ee
Thus, if ${\bf X}$ is a KV then ${\bf X} \cdot {\bf p}$ is a conserved quantity
along geodesics. If the CKV ${\bf X}$ is not a KV, then ${\bf X} \cdot {\bf p}$
is conserved along null geodesics only.

\section{Robertson-Walker Spacetimes}\label{sec:rwspacetimes}
The metric of a 3-dimensional space of constant curvature is 
\be 
ds^2 ={dr^2 \over 
(1-K r^2/{\cal R}^2)}+{r^2} d\Omega^2
\qquad \hbox{where} \qquad
d\Omega^2 = d\theta^2+\sin^2\theta d\phi^2 \: \: .
\label{eq:frwpolar} 
\ee 
where $K=+1, 0, -1$ indicating positive, zero or negative curvature 
respectively and ${\cal R}$ is the radius of curvature. 
We define $k=K/{\cal R}^2$ to be the curvature, that is, the parameter $k$ 
takes on {\bf all} values including zero in the metric
\be 
ds^2 ={dr^2 \over (1-k r^2)}+{r^2} \, d\Omega^2 \: \: .
\label{eq:frwpolargeneralk}
\ee
Consider the line element (\ref{eq:frwpolargeneralk}) and note that when 
$k=-1,0,+1$ we can write 
\be
ds^2 \: = \: d\chi^2 + r^2(\chi) \: d\Omega^2 
\label{eqn:rwmetric1}
\ee
where $d\chi = dr / {(1-kr^2)^{1 \over 2}}$. The function 
$r(\chi)$ is $\sin\chi$ for spherical spatial geometry, $\chi$ for flat 
spatial geometry and $\sinh\chi$ for hyperbolic spatial geometry. The line element for the general Robertson-Walker spacetime can be written 
\be
ds_{RW}^2=-dt^2+S^2(t) \, ds^2 \: \: .
\label{eqn:rwmetric2}
\ee
The function $S(t)$ is the cosmological scale factor.

\subsection{Conformally Flat Spacetime}
\subsubsection{Transformation to conformal time $\tau$}
First we write the general Robertson-Walker metric as 
\ba
ds_{RW}^2 & = & S^2(t) \, \biggl(-{{dt^2} \over {S^2(t)}} + ds^2 \biggr) \\
& = & S^2(t) \, (-d\tau^2 + ds^2) \qquad \hbox{where} \qquad
\tau = \int^{}_{} {dt \over {S(t)}} \: \: . 
\label{eqn:conftime}
\ea
The line element $ds_E^2 =-d\tau^2 + ds^2$ in the case $k>0$ is normally 
referred to as the Einstein static spacetime and in the case $k<0$, the 
anti-Einstein static spacetime. However, from now on we shall refer to the 
three cases $k>0$, $k=0$ and $k<0$ collectively as the 
{\it Einstein static spacetimes}.

\subsubsection{Einstein static spacetimes in conformally flat form}
We now present a coordinate transformation of the form
\be
ds_E^2 \: = \: C^2(\tau',r',k) \: ds^2_M (\tau',r',\theta', \phi')
\ee
where $C(\tau',r',k)$ is some function (to be determined) depending on the 
value of $k$ and $ds^2_M$ is the line element of Minkowski spacetime 
\be
ds_M^2 = -d\tau'^2 + dr'^2 + r'^2 (d\theta'^2 + \sin^2 \theta' d\phi'^2 )
\: \: .
\label{eqn:minkowski1} 
\ee
We first demonstrate that the Einstein static spacetime with $k=+1$ can be 
written in a conformally flat form (we do this by considering null coordinates 
in the Minkowski spacetime as described in appendix 
\ref{app:einsteinstatic:confflat}). 
Then we generalise this to all Einstein static spacetimes of arbitrary 
curvature $k$. 

For the case $k=+1$ the coordinate transformation is as follows 
\ba 
\tau' & = & {2 \sin \tau \over (\cos \tau + \cos \chi)} \: ; \nonumber\\
r' & = & {2 \sin \chi \over (\cos \tau + \cos \chi)} \: .
\label{eqn:tr2}
\ea
The line element (\ref{eqn:minkowski1}) becomes 
\be
ds^2_M = {4 \over (\cos \tau + \cos \chi)^2} \, 
(-d\tau^2 + d\chi^2 + \sin^2 \chi \, d\Omega^2) 
\label{eqn:minko6}
\ee
which is conformal to the Einstein static spacetime in the case where $k=+1$. 
We use the transformation (\ref{eqn:tr2}) and line element 
(\ref{eqn:minko6}) as our specimen. We generalise the transformation 
(\ref{eqn:tr2}) to accomodate spacetimes with arbitrary values of curvature 
$k$. 

Now, we can define a new time coordinate $\epsilon$ via
$d\tau = d\epsilon / {(1-k\epsilon^2)^{1 \over 2}}$ and radial coordinate $r$ 
via $d\chi = dr / {(1-kr^2)^{1 \over 2}}$ for $-\infty < k < \infty$. 
Note that when $k=+1,0,-1$ we have 
\be 
\epsilon = \left\{ \begin{array}{ll}
                   \sin \tau  & \mbox{$k=+1$} \\
                   \tau  & \mbox{$k=0$} \\
                   \sinh \tau  & \mbox{$k=-1$} \\
              \end{array}
            \right. ;
\qquad
r = \left\{ \begin{array}{ll}
                   \sin \chi  & \mbox{$k=+1$} \\
                   \chi  & \mbox{$k=0$} \\
                   \sinh \chi  & \mbox{$k=-1$} \\
              \end{array}
            \right.
\ee
Thus we have 
\be
ds_M^2 \: = \: {4 \over F^2} 
\biggl(-{{d\epsilon^2} \over {(1-k\epsilon^2)}} + 
{{dr^2} \over {(1-kr^2)}} + r^2 \; d\Omega^2 \biggr) 
\label{eqn:minkowski7}
\ee
where
\be
F(\tau,r)=(1-k\epsilon^2)^{1 \over 2} + (1-kr^2)^{1 \over 2} \: \: .
\ee
The coordinate transformation is 
\be
\tau' = {2 \; \epsilon \over {F(\tau,r)}} \, ; \qquad 
r' = {2 \; r \over {F(\tau,r)}} \: \: . \label{eq:transffrw1}
\ee
We note that
\be
F(\tau',r') = \: 4 \: {( \: {[(1+k\tau'^2 / 2) + 
(1+kr'^2 /2)]^2} 
- k^2 \tau'^2 r'^2 \; )^{-{1 \over 2}}} 
\ee
The inverse transformations to (\ref{eq:transffrw1}) are
\be
\epsilon = F(\tau',r') \, \tau'/2 \, ;\qquad 
r = F(\tau',r') \, r'/2 \: \: .
\ee
We note that, with the exception of the case $k=0$, our coordinate 
transformations are only valid locally. There does not exist a global 
coordinate transformation mapping all of Robertson-Walker spacetime 
conformally into Minkowski spacetime. 
However, the geometrical objects exist independently of any particular 
coordinate system and so we shall not mention these issues any further. 

\section{The CKV of Minkowski spacetime.} 
The line element for Minkowski spacetime (\ref{eqn:minkowski1}) can be written 
in cartesian coordinates 
\be
x'=r' \sin \theta' \cos \phi' \, , \qquad 
y'=r' \sin \theta' \sin \phi' \, , \qquad
z'=r' \cos \theta' \, ,
\ee
as follows
\be
ds_M^2 = -d\tau'^2 + dx'^2 + dy'^2 +dz'^2 \: \: . \label{eqn:minkowski2}
\ee
The isometry group of this spacetime is the 10-parameter poincare group (inhomogeneous Lorentz group) $ISO(3,1)$, the generators being the 6 
generators of the homogeneous Lorentz group $SO(3,1)$ and the 4 spacetime 
translations, giving 10 independent Killing vectors.

The conformal isometry group of Minkowski spacetime is the 15-parameter 
{\it conformal group} $SO(4,2)$ and includes the isometry group $ISO(3,1)$. 
The 15 CKV are 
\ba
\hbox{4 translations} & & \qquad {\bf T}_\alpha = 
{\partial \over {\partial x'^\alpha}}
\nonumber\\
\hbox{6 rotations} & & \qquad {\bf M}_{\alpha\beta} = 
x'_\alpha{\partial \over {\partial x'^\beta}}
-x'_\beta{\partial \over {\partial x'^\alpha}}
\nonumber\\
\hbox{1 dilation} & & \qquad {\bf D} = 
x'^\alpha{\partial \over {\partial x'^\alpha}}
\nonumber\\
\hbox{4 inversions} & & \qquad {\bf K}_\alpha = 
2x'_\alpha x'^\beta{\partial \over {\partial x'^\beta}}
-(x'^\beta x'_\beta){\partial \over {\partial x'^\alpha}}
\label{eqn:ckvminkowski1}
\ea
where $x'_\alpha=\eta_{\alpha\beta}x'^\beta$ and we shall refer to this as the 
Minkowski basis.
We can see from Table \ref{tab:minkowskickv} that with this choice of basis for 
the conformal algebra all the CKV are SCKV and there are only 4 proper CKV. 
The remaining 11 are HKV, and of course, only one 
of these is not a KV. The 15 CKV form a basis for the Lie algebra $so(4,2)$.
The commutation relations for these vector fields are as follows
\ba
& & [{\bf M}_{\alpha\beta}, {\bf M}_{\gamma\delta}] = 
\eta_{\alpha\delta}{\bf M}_{\beta\gamma} + 
\eta_{\beta\gamma}{\bf M}_{\alpha\delta} +
\eta_{\alpha\gamma}{\bf M}_{\delta\beta} +
\eta_{\beta\delta}{\bf M}_{\gamma\alpha}; \nonumber\\
& & [{\bf T}_\alpha, {\bf T}_\beta] = 0; \qquad
[{\bf T}_\alpha, {\bf M}_{\beta\gamma}] = 
\eta_{\beta\alpha}{\bf T}_\gamma - \eta_{\gamma\alpha}{\bf T}_\beta;\nonumber\\
& & [{\bf K}_\alpha, {\bf K}_{\beta}] = 0; \qquad \!
[{\bf K}_\alpha, {\bf M}_{\beta\gamma}]  =  
\eta_{\beta\alpha}{\bf K}_\gamma - \eta_{\gamma\alpha}{\bf K}_\beta;
\nonumber\\
& & [{\bf D}, {\bf K}_{\alpha}] = {\bf K}_{\alpha};\qquad
[{\bf D}, {\bf M}_{\alpha\beta}] = 0; \qquad 
[{\bf T}_\alpha, {\bf D}] = {\bf T}_\alpha;  
\nonumber\\
& & [{\bf T}_\alpha, {\bf K}_{\beta}] =  
2 (\eta_{\alpha\beta}{\bf D} - {\bf M}_{\alpha\beta}); 
\label{alg:min:iso31}
\ea
%
which is isomorphic to the Lie algebra (\ref{app:alg:so42:2}) presented in 
appendix \ref{app:conformalgroup}.
The subalgebra consisting of rotations ${\bf M}_{ij}$ and boosts ${\bf M}_{0k}$ is isomorphic to the Lie algebra $so(3,1)$. This subalgebra, together with the translations ${\bf T}_\alpha$ form the Lie algebra $iso(3,1)$.
It is obvious that the subalgebra of $so(4,2)$ consisting of rotations 
${\bf M}_{\alpha\beta}$ and the SCKV ${\bf K}_\alpha$ is also isomorphic to 
the Lie algebra $iso(3,1)$. The group of translations is an invariant 
subgroup of $ISO(3,1)$ and so the Lie group $ISO(3,1)$ is nonsemi-simple. 
The de Sitter and anti-de Sitter groups $SO(4,1)$ and $SO(3,2)$ are also 
subgroups of the conformal group $SO(4,2)$, however, they are not isometry 
groups of the Minkowski spacetime. Thus, the algebra $so(4,2)$ contains the 
subalgebras $so(4,1)$ and $so(3,2)$ which is shown explicitly in  
appendix \ref{app:conformalgroup} and section \ref{sec:homoisost}. 

In order to implement the transformation (\ref{eq:transffrw1}) to obtain the 
CKV for Robertson-Walker spacetimes we re-write the Minkowski CKV 
(\ref{eqn:ckvminkowski1}) in terms of spherical polar coordinates as follows 
\ba
{\bf T}_0 & = & {\partial \over {\partial \tau'}}
\nonumber\\
{\bf T}_1 & = & \sin\theta \cos\phi{\partial \over {\partial r'}}
+{1 \over r'}\biggl({\cos\theta \cos\phi}{\partial \over {\partial \theta}}
-{{\sin\phi} \over {\sin\theta}} {\partial \over {\partial \theta}} \biggr)
\phantom{00000000000000000}
\nonumber\\
{\bf T}_2 & = & \sin\theta \sin\phi{\partial \over {\partial r'}}
+{1 \over r'}\biggl({\cos\theta \sin\phi}{\partial \over {\partial \theta}}
+{{\cos\phi} \over {\sin\theta}} {\partial \over {\partial \theta}} \biggr)
\nonumber\\
{\bf T}_3 & = & \cos\theta{\partial \over {\partial r'}}
-{{\sin\theta} \over r'}{\partial \over {\partial \theta}}
\nonumber\\
{\bf M}_{01} & = & 
\sin\theta \cos\phi \biggl(-\tau'{\partial \over {\partial r'}}-
r'{\partial \over {\partial \tau'}}\biggr)
-{\tau' \over r'}\biggl({\cos\theta \cos\phi}{\partial \over {\partial \theta}}
-{{\sin\phi} \over {\sin\theta}} {\partial \over {\partial \phi}} \biggr)
\nonumber\\
{\bf M}_{02} & = & 
\sin\theta \sin\phi \biggl(-\tau'{\partial \over {\partial r'}}-
r'{\partial \over {\partial \tau'}}\biggr)
-{\tau' \over r'}\biggl({\cos\theta \sin\phi}{\partial \over {\partial \theta}}
+{{\cos\phi} \over {\sin\theta}} {\partial \over {\partial \phi}} \biggr)
\nonumber\\
{\bf M}_{03} & = &
\cos\theta \biggl(-\tau'{\partial \over {\partial r'}}-
r'{\partial \over {\partial \tau'}}\biggr)
+{{\tau'\sin\theta} \over r'}{\partial \over {\partial \theta}}
\nonumber\\
{\bf M}_{12} & = & 
{\partial \over {\partial \phi}}
\nonumber\\
{\bf M}_{13} & = & 
-\cos\phi{\partial \over {\partial \theta}}
+{{\cot\theta \sin\phi}} {\partial \over {\partial \phi}}
\nonumber\\
{\bf M}_{23} & = & 
-\sin\phi{\partial \over {\partial \theta}}
-{{\cot\theta \cos\phi}} {\partial \over {\partial \phi}}
\nonumber\\
{\bf D} & = & 
\tau'{\partial \over {\partial \tau'}}+r'{\partial \over {\partial r'}}
\nonumber\\
{\bf K}_0 & = & 
-2\tau' \; {\bf D}
-(-\tau'^2 +r'^2) \, {\bf T}_0
\nonumber\\
{\bf K}_1 & = & 
2r'\sin\theta \cos \phi \; {\bf D}
-(-\tau'^2 +r'^2) \, {\bf T}_1
\nonumber\\
{\bf K}_2 & = & 
2r'\sin\theta \sin \phi \; {\bf D}
-(-\tau'^2 +r'^2) \, {\bf T}_2
\nonumber\\
{\bf K}_3 & = & 
2r'\cos\theta \; {\bf D}
-(-\tau'^2 +r'^2) \, {\bf T}_3
\label{eqn:ckvminkowski2}
\ea

\section{The CKV of Robertson-Walker spacetimes} 
We now apply the transformation (\ref{eq:transffrw1}) to the Minkowski CKV 
(\ref{eqn:ckvminkowski2}). In doing this we will have simply re-written the 
Minkowski CKV in a different coordinate system so that the algebra remains 
unchanged. 
However we will also obtain a convenient basis for the Robertson-Walker CKV.   
From (\ref{eq:transffrw1}) we find that
\ba
{\partial \over {\partial \tau'}}
& = & {1 \over 2} \, \biggl(A {\partial \over {\partial \tau}} 
-k \epsilon r (1-kr^2)^{1 \over 2}
{\partial \over {\partial r}} \biggl)
\nonumber\\
{\partial \over {\partial r'}}
& = & 
{1 \over 2} \, \biggl(A (1-kr^2)^{1 \over 2} {\partial \over {\partial r}}
-k \epsilon r {\partial \over {\partial \tau}} \biggr)
\label{eqn:newbasisforfrw}
\ea
where $A=1+ (1-kr^2)^{1 \over 2}(1-k\epsilon^2)^{1 \over 2}$.
The relations (\ref{eqn:newbasisforfrw}) are essentially all we need to 
construct the CKV for the RW spacetimes. They are as follows
\ba
{\bf T}_0 & = & {1 \over 2} \, \biggl(A {\partial \over {\partial \tau}} 
-k \epsilon r (1-kr^2)^{1 \over 2}
{\partial \over {\partial r}} \biggl)
\nonumber\\
{\bf T}_1 & = & {1 \over 2} \, \sin\theta \cos\phi 
\biggl(A (1-kr^2)^{1 \over 2} {\partial \over {\partial r}}
-k \epsilon r {\partial \over {\partial \tau}} \biggr)
+{F \over 2r}\biggl({\cos\theta \cos\phi}{\partial \over {\partial \theta}}
-{{\sin\phi} \over {\sin\theta}} {\partial \over {\partial \phi}} \biggr)
\phantom{00000000000000000}
\nonumber\\
{\bf T}_2 & = & {1 \over 2} \, \sin\theta \sin\phi
\biggl(A (1-kr^2)^{1 \over 2} {\partial \over {\partial r}}
-k \epsilon r {\partial \over {\partial \tau}} \biggr)
+{F \over 2r}\biggl({\cos\theta \sin\phi}{\partial \over {\partial \theta}}
+{{\cos\phi} \over {\sin\theta}} {\partial \over {\partial \phi}} \biggr)
\nonumber\\
{\bf T}_3 & = & {1 \over 2} \, \cos\theta
\biggl(A (1-kr^2)^{1 \over 2} {\partial \over {\partial r}}
-k \epsilon r {\partial \over {\partial \tau}} \biggr)
-{F \over 2r} \, {{\sin\theta}}{\partial \over {\partial \theta}}
\nonumber\\
{\bf M}_{01} & = & 
-\sin\theta \cos\phi 
\biggl((1-k\epsilon^2)^{1 \over 2} r {\partial \over {\partial \tau}} + 
(1-kr^2) \epsilon {\partial \over {\partial r}} \biggr)
-{\epsilon \over r}\biggl({\cos\theta \cos\phi}{\partial \over {\partial \theta}}
-{{\sin\phi} \over {\sin\theta}} {\partial \over {\partial \phi}} \biggr)
\nonumber\\
{\bf M}_{02} & = & 
-\sin\theta \sin\phi 
\biggl( (1-k\epsilon^2)^{1 \over 2} r {\partial \over {\partial \tau}} + 
(1-kr^2) \epsilon {\partial \over {\partial r}} \biggr)
-{\epsilon \over r}\biggl({\cos\theta \sin\phi}{\partial \over {\partial \theta}}
+{{\cos\phi} \over {\sin\theta}} {\partial \over {\partial \phi}} \biggr)
\nonumber\\
{\bf M}_{03} & = &
-\cos\theta 
\biggl( (1-k\epsilon^2)^{1 \over 2} r {\partial \over {\partial \tau}} + (1-kr^2) \epsilon {\partial \over {\partial r}} \biggr)
+{\epsilon \over r} \sin\theta {\partial \over {\partial \theta}}
\nonumber\\
{\bf M}_{12} & = & 
{\partial \over {\partial \phi}}
\nonumber\\
{\bf M}_{13} & = & 
-\cos\phi{\partial \over {\partial \theta}}
+{{\cot\theta \sin\phi}} {\partial \over {\partial \phi}}
\nonumber\\
{\bf M}_{23} & = & 
-\sin\phi{\partial \over {\partial \theta}}
-{{\cot\theta \cos\phi}} {\partial \over {\partial \phi}}
\nonumber\\
{\bf D} & = & (1-kr^2)^{1 \over 2} \biggl(
\epsilon \,  {\partial \over {\partial \tau}}
+ (1-k\epsilon^2)^{1 \over 2} r
{\partial \over {\partial r}} \biggr)
\nonumber\\
{\bf K}_0 & = & 
-{4\epsilon \over F} \; {\bf D} 
-{4 \over F^2}(-\epsilon^2 +r^2) \, {\bf T}_0
\nonumber\\
{\bf K}_1 & = & 
{4r \over F}\sin\theta \cos \phi  \; {\bf D}
-{4 \over F^2}(-\epsilon^2 +r^2) \, {\bf T}_1
\nonumber\\
{\bf K}_2 & = & 
{4r \over F}\sin\theta \sin \phi  \; {\bf D}
-{4 \over F^2}(-\epsilon^2 +r^2) \, {\bf T}_2
\nonumber\\
{\bf K}_3 & = & 
{4r \over F}\cos\theta  \; {\bf D}
-{4 \over F^2}(-\epsilon^2 +r^2) \, {\bf T}_3
\label{eqn:ckvrw}
\ea
The rotations ${\bf M}_{ij}$ are obviously invariant under the map.
Of course, the commutation relations are still as in the Lie algebra 
(\ref{alg:min:iso31}).
These CKV can be expressed in terms of the pseudo-isotropic coordinates 
(appendix \ref{app:conflatspatial}) and can be written as
\ba
{\bf T}_0 & = & {1 \over 2} \, ({\bf P}_0 + {\bf H} ) \: ; \nonumber\\
{\bf T}_i & = & {1 \over 2} \, ({\bf P}_i + {\bf {\bar Q}}^*_i ) 
\phantom{00000000000000000000000000000000000000000000000000000}
\label{eqn:ckvfrwnewbasis1}
\ea
where 
\ba
{\bf P}_0 & = & {\partial \over \partial \tau} \nonumber\\
{\bf P}_i & = & K_- {\partial \over \partial {\bar x}^i} 
+ {k \over 2} {\bar x}^i 
\biggl( {\bar x}^j {\partial \over \partial {\bar x}^j} \biggr)
\nonumber\\
{\bf H} & = & 
K_-K_+^{-1} (1-k \epsilon^2)^{1 \over 2} {\partial \over \partial \tau}
- k \epsilon \biggl( {\bar x}^j {\partial \over \partial {\bar x}^j} \biggr)
\nonumber\\
{\bf {\bar Q}}^*_i & = & -k \epsilon K_+^{-1} {\bar x}^i {\partial \over \partial \tau} + (1-k \epsilon^2)^{1 \over 2} 
\biggl( -{\bf P}_i + 2 {\partial \over \partial {\bar x}^i} \biggr)
\label{eqn:ckvrw2}
\ea
and 
\ba
{\bf M}_{0i} & = & -(1-k \epsilon^2)^{1 \over 2} K_+^{-1} {\bar x}^i 
{\partial \over \partial \tau}
-\epsilon \biggl( -{\bf P}_i + 2 {\partial \over \partial {\bar x}^i} \biggr) \nonumber\\
{\bf M}_{ij} & = & {\bar x}^i {\partial \over \partial {\bar x}^j} - 
{\bar x}^j {\partial \over \partial {\bar x}^i} 
\nonumber\\
{\bf D} & = & \epsilon  \, K_-K_+^{-1} {\partial \over \partial \tau}
+ (1-k \epsilon^2)^{1 \over 2} \biggl( {\bar x}^j 
{\partial \over \partial {\bar x}^j} \biggr) \nonumber\\
{\bf K}_0 & = & -{4 \epsilon \over ((1-k \epsilon^2)^{1 \over 2} + 
K_-K_+^{-1})} \, {\bf D} - {4 \over ((1-k \epsilon^2)^{1 \over 2} + 
K_-K_+^{-1})^2} (-\epsilon^2+ {\bar r}^2 k_+^{-2}) \, {\bf T}_0
\nonumber\\
{\bf K}_i & = & {4 K_+^{-1} {\bar x}^i \over ((1-k \epsilon^2)^{1 \over 2} + 
K_-K_+^{-1})} \, {\bf D} - {4 \over ((1-k \epsilon^2)^{1 \over 2} + 
K_-K_+^{-1})^2} (-\epsilon^2+ {\bar r}^2 k_+^{-2}) \, {\bf T}_i
\label{eqn:ckvrw3}
\ea
The conformal factors for the CKV (\ref{eqn:ckvrw}) are presented in Table 
\ref{tab:frwckv}.
Thus we can see that, with the exception of the rotations ${\bf M}_{ij}$, the 
SCKV of Minkowski spacetime are mapped into proper CKV in the Robertson-Walker 
spacetimes. Only in special cases will some of these proper CKV reduce to SCKV, 
see sections \ref{sec:contbasis} and \ref{sec:homoisost}.
Of course, the structure constants in the Lie algebra (\ref{alg:min:iso31}) 
do not involve the curvature parameter $k$. 
Thus if we wish to recover, say, the 6-dimensional isometry subalgebras 
($so(4), iso(3)$, or $so(3,1)$ for positive, zero and negative values of $k$)  whose structure constants depend upon the value of $k$, it will be necessary to make a change of basis. The form of the CKV (\ref{eqn:ckvfrwnewbasis1}) is 
very suggestive as it is well known that the vector fields ${\bf P}_i$ are 
Killing vector fields.
In the following sections we will construct such bases.

\subsection{The basis of Maartens and Maharaj}
\begin{figure*}[here!]
 \begin{center}{
   \epsfxsize 10.0 true cm
    \leavevmode
 \rotatebox{0}{\epsffile{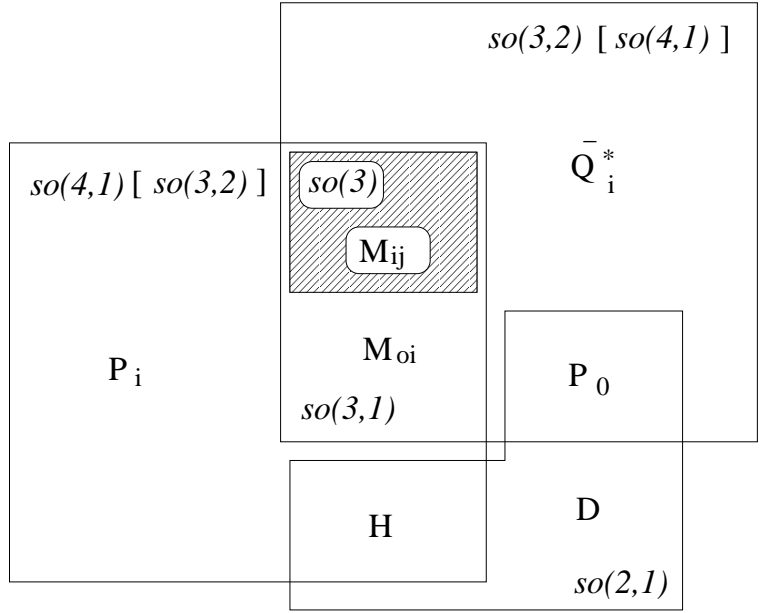}}
  }\end{center}\caption[]{\small{The subalgebra structure is shown 
schematically for the basis of Maartens and Maharaj 
(\ref{eqn:ckvfrwnewbasis:desitter}).
The algebras appearing in square brackets apply in the case $k<0$.}}
\label{fig:frw1}
\end{figure*}

Now, it can be shown that 
\ba
k \, {\bf K}_0 & = & 2 \, (-{\bf P}_0 + {\bf H})  \nonumber\\
k \, {\bf K}_i & = & 2 \, ({\bf P}_i - {\bf {\bar Q}}^*_i) \: \: .
\label{eqn:ckvfrwnewbasis2}
\ea
Taking into account the relations (\ref{eqn:ckvfrwnewbasis1}) and (\ref{eqn:ckvfrwnewbasis2}) we can 
write 
\ba
{\bf P}_0 & = & {\bf T}_0 - {k \over 4} \, {\bf K}_0 \nonumber\\
{\bf H} & = & {\bf T}_0 + {k \over 4} \, {\bf K}_0 \nonumber\\
{\bf P}_i & = & {\bf T}_i + {k \over 4} \, {\bf K}_i \nonumber\\
{\bf {\bar Q}}^*_i & = & {\bf T}_i - {k \over 4} \, {\bf K}_i 
\label{eqn:ckvfrwnewbasis3}
\ea
The conformal factors for these CKV are shown in Table 
\ref{tab:frwckv-newbasis}.
Let us consider the basis
\be
{\bf P}_0; \: {\bf P}_i;  \: {\bf H};  \: 
{\bf {\bar Q}}^*_i;  \: {\bf M}_{0i};  \: {\bf M}_{ij};  \: {\bf D};
\label{eqn:ckvfrwnewbasis:desitter}
\ee 
as an alternative basis for the Lie algebra $so(4,2)$.
Now, it can be easily seen that when $k=0$ we have ${\bf P}_0={\bf H}$ and 
${\bf P}_i={\bf {\bar Q}}^*_i$. Thus when $k=0$ the vector fields 
(\ref{eqn:ckvfrwnewbasis:desitter}) will not span the 15-dimensional vector 
space. Thus the basis (\ref{eqn:ckvfrwnewbasis:desitter}) is only legitimate in 
the case where $k \ne 0$. However, we shall consider this basis further.
Table \ref{tab:frwckv-newbasis-algebra} shows the structure of the 
the Lie algebra in terms of this basis.
The subalgebra formed by the elements
$\{ {\bf H}; \:{\bf P}_i; \: {\bf M}_{0i}; \: {\bf M}_{ij} \}$
and the subalgebra formed by the elements
$\{ {\bf P}_0; \: {\bf {\bar Q}}^*_i;  \: {\bf M}_{0i}; \: {\bf M}_{ij} \}$
are complimentary in the sense that when we scale out $k$, these have the 
structure $so(4,1)$ and $so(3,2)$ respectively for the case $k>0$ and 
$so(3,2)$ and $so(4,1)$ respectively when $k<0$. 
The subalgebra $\{ {\bf M}_{0i}; \: {\bf M}_{ij} \}$ has the structure 
$so(3,1)$.
The subalgebra formed by $\{ {\bf P}_i, {\bf M}_{jk} \}$ has the structure 
$so(4)$ or $so(3,1)$ according to whether $k>0$ or $k<0$ and the subalgebra 
formed by $\{ {\bf {\bar Q}}^*_i, {\bf M}_{jk} \}$ has the structure 
$so(3,1)$ or $so(4)$ according to whether $k>0$ or $k<0$.
Of course, the former is the KV algebra of the spacetime. 
The subalgebra $\{ {\bf P}_0, {\bf D}, {\bf H} \}$ has the structure constants 
of the Lie algebra $so(2,1)$. The subalgebra structure is shown schematically 
in Figure \ref{fig:frw1}.
Let us consider the CKV in the case where $k= \pm 1$. We find 
that the CKV presented above are related to those in 
Maartens and Maharaj [3] in this case by the following 
\ba
& & {\bf {\bar Q}}^*_i = -k \, {\bf Q}^*_i \nonumber\\
& & {\bf M}_{0i} = k \, {\bf Q}_i \nonumber\\
& & {\bf D} = -k \, {\bf H}^*  \: \: .
\ea
where they have $h(\tau)=(1-k\epsilon^2)^{1 \over 2}$.
Thus the Lie algebra for the cases $k=\pm 1$ can be written as follows
\ba
& & [{\bf P}_0, {\bf P}_i] = 0; \qquad 
[{\bf P}_0, {\bf M}_{ij}] = 0; \qquad 
[{\bf P}_i, {\bf P}_j] = -k \, {\bf M}_{ij}; \nonumber\\ 
& & [{\bf M}_{ij}, {\bf M}_{mn}]  =  
\eta_{jm}{\bf M}_{in} + 
\eta_{jn}{\bf M}_{mi} +
\eta_{im}{\bf M}_{nj} +
\eta_{in}{\bf M}_{jm}; \nonumber\\ 
& & [{\bf P}_i, {\bf M}_{jk}] = \eta_{ij}{\bf P}_k - \eta_{ik}{\bf P}_j;
\qquad 
[{\bf P}_i, {\bf H}] = {\bf Q}_i; \qquad 
[{\bf P}_i, {\bf H}^*] = {\bf Q}^*_i; \nonumber\\ 
& & [{\bf P}_0, {\bf H}] = {\bf H}^*; \qquad
[{\bf P}_0, {\bf H}^*] = -k \, {\bf H}; \qquad
[{\bf P}_0, {\bf Q}_i] = {\bf Q}^*_i; \qquad
[{\bf P}_0, {\bf Q}^*_i] = -k \, {\bf Q}_i; \nonumber\\
& & [{\bf P}_i, {\bf Q}_j] = -k \, \eta_{ij} {\bf H}; \qquad 
[{\bf P}_i, {\bf Q}^*_j]  =  -k \, \eta_{ij} {\bf H}^*; \qquad 
[{\bf M}_{ij}, {\bf H}] = 0; \qquad 
[{\bf M}_{ij}, {\bf H}^*] = 0; \nonumber\\
& & [{\bf Q}_i, {\bf M}_{jk}] = \eta_{ij} {\bf Q}_k - \eta_{ik} {\bf Q}_j; 
\qquad 
[{\bf Q}^*_i, {\bf M}_{jk}] = \eta_{ij} {\bf Q}^*_k - \eta_{ik} {\bf Q}^*_j; 
\nonumber\\ 
& & [{\bf H}, {\bf H}^*] = -k {\bf P}_0 ; \qquad 
[{\bf Q}^*_i,{\bf H}] = 0; \qquad 
[{\bf Q}_i,{\bf H}^*] = 0; \qquad 
[{\bf Q}_i, {\bf H}] = k \, {\bf P}_i; \nonumber\\
& & [{\bf Q}^*_i,{\bf H}^*] = {\bf P}_i; \qquad 
[{\bf Q}_i,{\bf Q}_j] = {\bf M}_{ij}; \qquad 
[{\bf Q}^*_i, {\bf Q}^*_j] = k \, {\bf M}_{ij};
\qquad 
[{\bf Q}_i, {\bf Q}^*_j] = - \eta_{ij} \, {\bf P}_0 \: .
\label{eqn:mmliealgebra}
\ea 
It should be noted that some of the commutation relations in the Lie algebra 
(3.4) presented in Maartens and Maharaj [3] are incorrect - 
the commutation relations 
$[{\bf M}_{ij},{\bf M}_{mn}]$, $[{\bf H}, {\bf H}^*]$,
$[{\bf Q}_i, {\bf H}]$ and $[{\bf Q}^*_i, {\bf Q}^*_j]$ are incorrect
and the relation $[{\bf Q}_i, {\bf Q}^*_j]$ is missing. 
We have rectified these in the Lie algebra (\ref{eqn:mmliealgebra}) above.

\subsection{The ``continuous" basis}\label{sec:contbasis}
\begin{figure*}[here!]
 \begin{center}{
   \epsfxsize 10.0 true cm
    \leavevmode
 \rotatebox{0}{\epsffile{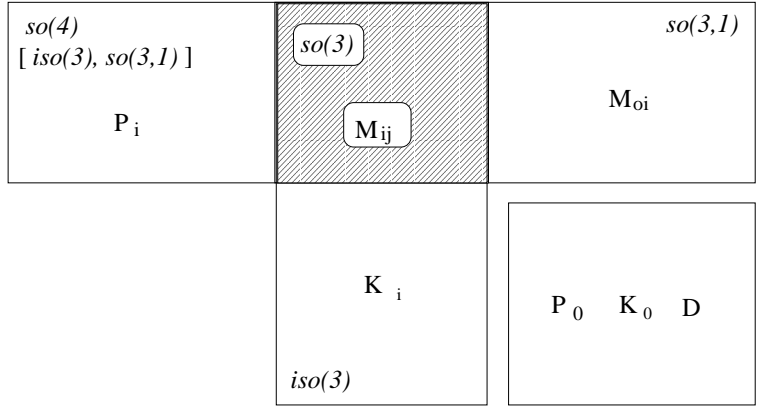}}
  }\end{center}\caption[]{\small{The subalgebra structure is shown 
schematically for the continuous basis (\ref{eqn:ckvfrwnewbasis:continuous}).
The algebras appearing in square brackets apply in the cases $k=0$ and $k<0$
respectively.}}
\label{fig:frw2}
\end{figure*}

Alternatively, we can choose 
\be
{\bf P}_0; \: {\bf P}_i;  \: {\bf M}_{0i};  \: {\bf M}_{ij};  \: {\bf D};
\: {\bf K}_{0};  \: {\bf K}_{i}; 
\label{eqn:ckvfrwnewbasis:continuous}
\ee
as a basis for the algebra $so(4,2)$. Again the full KV algebra is 
a subalgebra of the complete Lie algebra. However, in this case the choice 
of basis is valid for {\it all} values of curvature parameter $k$ including the 
case $k=0$. The Lie algebra is shown in Table \ref{tab:banana}.
The subalgebra formed by the elements $\{ {\bf P}_i, {\bf M}_{jk} \}$ 
corresponds to the KV subalgebras $so(4)$, $iso(3)$ or $so(3,1)$ for $k>0$, 
$k=0$ and $k<0$ respectively.
The subalgebra formed by the elements $\{ {\bf M}_{0i}, {\bf M}_{jk} \}$
corresponds to the Lie algebra $so(3,1)$.
The subalgebra formed by $\{ {\bf K}_i, {\bf M}_{jk} \}$
corresponds to the algebra $iso(3)$. However, ${\bf M}_{0i}$ is not a KV in 
general and the ${\bf K}_i$ are never KV.
The subalgebra structure is shown schematically in Figure \ref{fig:frw2}.

${\bf P}_0$ becomes a HKV when $S(t)=Ct$ where $C=constant$ and becomes 
a KV when $S(t)=C$ and in this case we recover the Einstein static spacetimes. 
However, in the case when $k=0$ and $S(t)=C$ the 
${\bf M}_{0i}$ are necessarily KV also and so there are an extra 4 KV. 
See Maartens and Maharaj [3] for further discussion. 


\section{De Sitter and anti-de Sitter spacetimes}\label{sec:homoisost}
Let us now consider spacetimes with constant curvature. These are a special 
case of the Robertson-Walker spacetimes. In this case the Riemann curvature 
tensor is determined by the Ricci scalar alone. These spacetimes are Einstein 
spaces i.e.
\be
 R_{\alpha\beta}= {1 \over 4} \, R \, g_{\alpha\beta} \: \: .
\ee
The spacetime of constant curvature with $R=0$ is Minkowski spacetime, $R>0$ is 
{\it de Sitter spacetime} and in the case $R<0$, {\it anti-de Sitter spacetime}. Of course, the spacetimes of constant curvature are maximally symmetric 
i.e. admit 10 independent Killing vector fields. Thus they are homogeneous and 
isotropic spacetimes. The generic de Sitter spacetime can be written 
(see Hawking and Ellis [7])
\be
ds^2 = -dt^2 + (1+R \sigma^2)
\biggl( {dr^2 \over (1-R r^2)} + r^2 d\Omega^2 \biggr)
\ee
for arbitrary values of curvature $R$, where 
$dt/d\sigma = 1/(1+R\sigma^2)^{1 \over 2}$.
It is obvious that this metric takes on the same form as that of a 
Robertson-Walker spacetime with $S(t)=(1+R\sigma^2)^{1 \over 2}$. 
However, these coordinates are distinct from those used in section 
\ref{sec:rwspacetimes} to describe the Robertson-Walker spacetimes.
We note that the constants $k$ and $R$ are related by the following 
\be 
R = 6 S^{-2} (S \, {\partial^2 S / \partial t^2} + 
({\partial S / \partial t})^2 + k) \: .
\label{eqn:Randk}
\ee
The correspondence noted above allows us to read off the CKV for the de Sitter 
family of spacetimes. Of course the 6 KV ${\bf P}_i, {\bf M}_{jk}$ will be 
inherited and in addition the 4 CKV ${\bf H}$ and ${\bf M}_{0i}$ are also KV 
for the de Sitter spacetimes. The quantity 
\be
\Phi= S^{-1} {\partial S / \partial \tau} 
(1-R\epsilon^2)^{1 \over 2}-R \epsilon
\ee
is zero for all $R$ and as a result the conformal factors for 
${\bf H}$ and ${\bf M}_{0i}$ become zero, see Tables 
\ref{tab:frwckv} and \ref{tab:frwckv-newbasis}.
It is easily verified that $\Phi=0$ for $R=0$.
For $R>0$ we can scale out $|R|$ and write $S=\cosh t$ and 
from (\ref{eqn:conftime}), $\tan \tau=\sinh t$.
We can also write $(1-R\epsilon^2)^{1 \over 2}=\cos \tau$ and 
$-R\epsilon=-\sin \tau$ and it follows that $\Phi=0$.
Similarly, for $R<0$ we can scale out $|R|$ and write $S=\cos t$ and 
from (\ref{eqn:conftime}), $\cosh \tau=\sec t$.
We can also write $(1-R\epsilon^2)^{1 \over 2}=\cosh \tau$ and 
$-R\epsilon=\sinh \tau$ and it follows that $\Phi=0$.

\begin{figure*}[here!]
 \begin{center}{
   \epsfxsize 10.0 true cm
    \leavevmode
 \rotatebox{0}{\epsffile{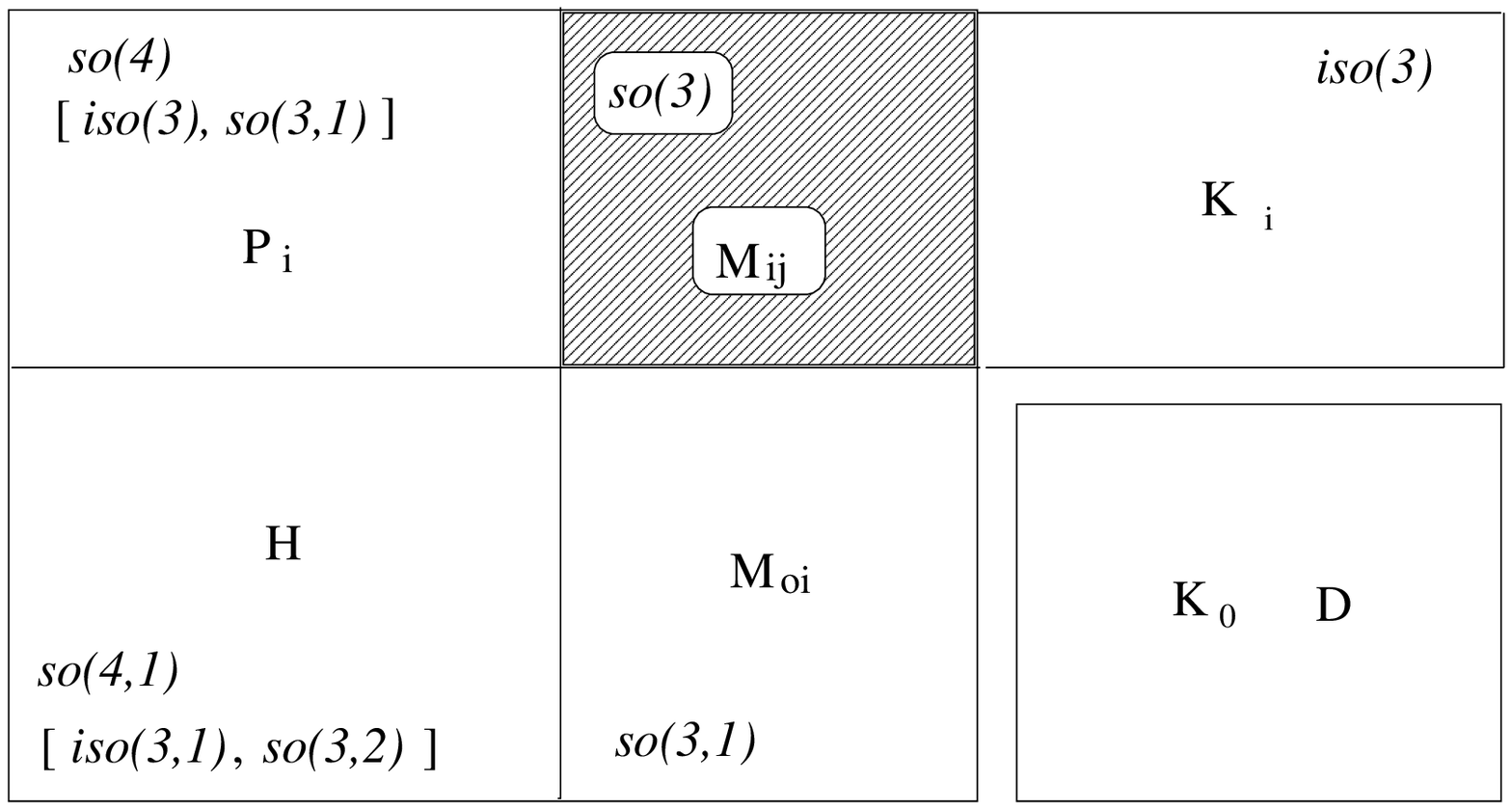}}
  }\end{center}\caption[]{\small{The subalgebra structure is shown 
schematically for the de Sitter basis
(\ref{eqn:ckvfrwnewbasis:continuous:desitter}).
The 10 vector fields 
${\bf H}; \: {\bf P}_i;  \: {\bf M}_{0i};  \: {\bf M}_{ij}$ 
form the KV subalgebra of the de Sitter basis. The algebras appearing in 
square brackets apply in the cases $R=0$ and $R<0$ respectively.}}
\label{fig:frw3}
\end{figure*}

Thus we now choose 
\be
{\bf H}; \: {\bf P}_i;  \: {\bf M}_{0i};  \: {\bf M}_{ij};  \: {\bf D};
\: {\bf K}_{0};  \: {\bf K}_{i}; 
\label{eqn:ckvfrwnewbasis:continuous:desitter}
\ee
as a basis for the algebra $so(4,2)$ and we shall refer to this as the 
``de Sitter" basis. The full KV algebra is a subalgebra of the 
complete algebra for {\it all} values of curvature parameter $R$ including the 
case $R=0$. The Lie algebra is shown in Table \ref{tab:frwckv-desitter} and 
schematically in Figure \ref{fig:frw3}.

The KV subalgebra 
$\{ {\bf H}; \: {\bf P}_i;  \: {\bf M}_{0i};  \: {\bf M}_{ij}\}$ 
is the general de Sitter algebra i.e. $so(4,1)$, $iso(3,1)$ 
or $so(3,2)$ according to whether $R>0$, $R=0$ or $R<0$.
The subalgebra of the de Sitter algebra with elements 
$\{ {\bf M}_{0i};  \: {\bf M}_{ij} \}$ form the Lie algebra $so(3,1)$ 
thus de Sitter spacetime will always possess a Lorentz (isometry) subgroup. 
The subalgebra of $so(4,2)$ with elements 
$\{ {\bf K}_i;  \: {\bf M}_{ij} \}$ forms the Lie algebra $iso(3)$. 

Of course, the de Sitter family of spacetimes can also be written in terms of 
the curvature parameter $k$. It can easily be seen from equation 
(\ref{eqn:Randk}) that the sign of $k$ is constrained by the sign of $R$.
In the case $R>0$ there are three distinct forms 
corresponding to $k>0$, $k=0$ and $k<0$, and for $R=0$ there are two forms 
given by $k=0$ and $k<0$. For the case $R<0$ there is only one possibility,
$k<0$, see Torrence and Couch [8].
Maartens and Maharaj [3] list all possible forms for the scale factors 
$S(t)$ for values of curvature $k$ in the Robertson-Walker models 
(\ref{eqn:rwmetric2}) in order that the spacetime be of constant curvature. 
All of these cases are encapsulated in the formalism presented above.

\section{Conclusions}
The mapping between Minkowski spacetime and the Robertson-Walker 
spacetimes provides a convenient way of analysing their conformal structure.
In particular the structure of the Lie algebra of conformal Killing vector 
fields is preserved by the mapping.
This has allowed us to make informed decisions regarding the choice of 
basis in particular situations of interest. 
For both the Robertson-Walker and de Sitter models we have presented 
bases which always contain the Killing vector basis and have shown 
that the Lie algebra structure can be continuously parameterised by the 
curvature $k$ (or the Ricci scalar $R$).

Of course, this type of analysis is valid for any spacetime which can be 
written in a manifestly conformally flat form although the bases presented 
here will not necessarily contain the Killing vector basis (if any).

\section*{Acknowledgements}
The authors would like to thank John Simmons for many useful discussions 
relating to symmetries and general relativity. We would also 
like to thank Graham Hall for useful comments on the first draft of this paper. 

\appendix
\section{Einstein static spacetimes in conformally flat form}
\label{app:einsteinstatic:confflat}
We can write the Einstein static spacetime in a conformally flat form by 
considering null coordinates in the Minkowski spacetime as described in 
Hawking and Ellis [7]. 

We define the {\it advanced and retarded null coordinates} $v$ and $w$ as follows
\be
v = {1 \over 4} (\tau'+r'); \qquad w = {1 \over 4} (\tau'-r') \: \: . 
\ee 
Note that Hawking and Ellis [7] omit the factor $1/4$ in their definition. 
The Minkowski line element (\ref{eqn:minkowski1}) becomes 
\be
ds^2_M = 16 \, (-dvdw + {1 \over 4} (v-w)^2 d\Omega^2) 
\label{eqn:minkowski3}
\ee
where $-\infty < v < \infty$, $-\infty < v < \infty$ and $v \ge w$. 
We can then define new null coordinates in which $v$ and $w$ are transformed to 
finite values i.e. $\tan p = v, \tan q = w$. Thus $-\pi/2<p<\pi/2$, 
$-\pi/2<q<\pi/2$ and $p \ge q$. The line element (\ref{eqn:minkowski3}) becomes
\be
ds^2_M = 16 \, \sec^2 \! p \; \sec^2 \! q \, 
(-dpdq + {1 \over 4} \sin^2(p-q) \, d\Omega^2) \: \: .
\label{eqn:minkowski4}
\ee
Define $\tau=p+q$, $\chi=p-q$ where $-\pi<\tau+\chi<\pi$,
 $-\pi<\tau-\chi<\pi$, 
$\chi \ge 0$. Finally, (\ref{eqn:minkowski4}) becomes
\be
ds^2_M = 4 \, \sec^2 \biggl({\tau+\chi \over 2} \biggr) \sec^2 
\biggl({\tau-\chi \over 2} \biggr) \, 
(-d\tau^2 + d\chi^2 + \sin^2 \chi \, d\Omega^2)
\label{eqn:minkowski5}
\ee
and the coordinates are related by 
\be
\tau' = 2 \tan \biggl({\tau+\chi \over 2} \biggr) + 
2 \tan \biggl({\tau-\chi \over 2} \biggr); \qquad 
r' = 2 \tan \biggl({\tau+\chi \over 2} \biggr) - 
2 \tan \biggl({\tau-\chi \over 2} \biggr) \: \: .
\label{eqn:transf1}
\ee
Note that the coordinate transformation (\ref{eqn:transf1}) can be 
re-expressed as 
\ba 
\tau' & = & 2\sin \tau \sec \biggl({\tau+\chi \over 2} \biggr) 
\sec \biggl({\tau-\chi \over 2} \biggr) = 
{2 \sin \tau \over (\cos \tau + \cos \chi)}; \nonumber\\
r' & = & 2\sin \chi \sec \biggl({\tau+\chi \over 2} \biggr) 
\sec \biggl({\tau-\chi \over 2} \biggr) = 
{2 \sin \chi \over (\cos \tau + \cos \chi)} \: \: .
\label{eqn:transf2}
\ea
This is the transformation presented in section 25.4 of Stephani [9].
The line element (\ref{eqn:minkowski5}) becomes 
\be
ds^2_M = {4 \over (\cos \tau + \cos \chi)^2} \, 
(-d\tau^2 + d\chi^2 + \sin^2 \chi \, d\Omega^2) 
\label{eqn:minkowski6}
\ee
and we recognise this as being conformal to the Einstein static spacetime in the case where $k=+1$.

\section{Conformally flat spatial sections}\label{app:conflatspatial}
We can use the transformation 
\be 
{\bar r}= {2r \over {1+(1-kr^2)^{1 \over 2}}}
\qquad \hbox{with inverse} \qquad
r={\bar r}  K_+^{-1}
\ee
where $K_{\pm} =1 \pm k {\bar r}^2 /4$. This transformation is valid for all 
values of k.
The spatial sections of the Robertson-Walker spacetimes   
(\ref{eq:frwpolargeneralk}) then take on the conformally flat form
\be
ds^2= K_+^{-2} (d{\bar r}^2 + {\bar r}^2 d\Omega^2) \: \: .
\ee
Further, defining pseudo-cartesian coordinates $\{ {\bar x}^i \}$ such that 
${\bar x}= {\bar r} \sin\theta \cos\phi$, 
${\bar y}= {\bar r} \sin\theta \sin\phi$,
${\bar z}= {\bar r} \cos\theta$, we have
\be
ds^2= K_+^{-2} (d{\bar x}^2 + d{\bar y}^2 + d{\bar z}^2)
\ee
where ${\bar r}=({\bar x}^2+{\bar y}^2+{\bar z}^2)^{1 \over 2}$. 

\section{Conformal group $SO(4,2)$}\label{app:conformalgroup}
The {orthogonal group} $O(p,q)$ leaves invariant the metric ${\bf \eta}$ on 
$I \! \! R^n$ with signature  
\be
( \: {\underbrace{ \: + \: + \: \dots \: }_p} \: ; \:
{\underbrace { \: - \: - \: \dots \: }_q} \: )
\ee
where $p+q=n$. The {\it special orthogonal group} $SO(p,q)$ is a Lie group 
which has group elements $g$ satisfying $det(g)=1$.
The associated Lie algebra, which we shall denote $so(p,q)$, has Killing 
vector generators 
\be
{\bf L}_{AB} = x_A {\partial \over \partial x^B} - 
x_B {\partial \over \partial x^A} \: \: , \label{eqn:confgroupbasis1}
\ee
where $A,B=1,2, \dots n$ and $x_A=\eta_{AB}x^B$. The Lie algebra has the form 
\ba
& & [{\bf L}_{AB}, {\bf L}_{CD}] = 
\eta_{AD}{\bf L}_{BC} + 
\eta_{BC}{\bf L}_{AD} +
\eta_{AC}{\bf L}_{DB} +
\eta_{BD}{\bf L}_{CA} \: \: . \label{app:alg:so42:1}
\ea
In particular the Lie group $SO(4,2)$ is the isometry group of the flat space 
with metric signature $(\: + \: + \: + \: - \: + \: - \: )$ and leaves 
Minkowski spacetime, with metric $(\: - \: + \: + \: + \:)$, conformally 
invariant. This Lie group is called the {\it conformal group}.  
We can choose the following basis 
\ba
& & {\bf M}_{\alpha\beta} = {\bf L}_{\alpha\beta}; \nonumber\\
& & {\bf T}_\alpha = {\bf L}_{\alpha 5} + {\bf L}_{\alpha 6}; \nonumber\\
& & {\bf K}_\alpha = {\bf L}_{\alpha 5} - {\bf L}_{\alpha 6}; \nonumber\\
& & {\bf D} = {\bf L}_{56} \label{eqn:confgroupbasis2}
\ea
where $\alpha = 0,1,2,3$ and $\alpha = 0 \equiv A=4$. We can write the 
Lie algebra in terms of the basis (\ref{eqn:confgroupbasis2}) as  
\ba
& & [{\bf M}_{\alpha\beta}, {\bf M}_{\gamma\delta}] = 
\eta_{\alpha\delta}{\bf M}_{\beta\gamma} + 
\eta_{\beta\gamma}{\bf M}_{\alpha\delta} +
\eta_{\alpha\gamma}{\bf M}_{\delta\beta} +
\eta_{\beta\delta}{\bf M}_{\gamma\alpha}; \nonumber\\
& & [{\bf T}_\alpha, {\bf T}_\beta] = 0; \qquad
[{\bf T}_\alpha, {\bf M}_{\beta\gamma}] = 
\eta_{\beta\alpha}{\bf T}_\gamma - \eta_{\gamma\alpha}{\bf T}_\beta;
\nonumber\\ 
& & [{\bf K}_\alpha, {\bf K}_{\beta}] = 0; \qquad 
[{\bf K}_\alpha, {\bf M}_{\beta\gamma}]  =  
\eta_{\beta\alpha}{\bf K}_\gamma - \eta_{\gamma\alpha}{\bf K}_\beta;
\nonumber\\
& & [{\bf D}, {\bf K}_{\alpha}] = {\bf K}_{\alpha};\qquad
[{\bf D}, {\bf M}_{\alpha\beta}] = 0; \qquad 
[{\bf T}_\alpha, {\bf D}] = {\bf T}_\alpha;  
\nonumber\\
& & [{\bf T}_\alpha, {\bf K}_{\beta}] =  
2 (\eta_{\alpha\beta}{\bf D} - {\bf M}_{\alpha\beta}). 
\label{app:alg:so42:2}
\ea
The Poincare group $ISO(3,1)$ is a subgroup of the conformal group $SO(4,2)$. 
In the basis (\ref{eqn:confgroupbasis2}) it is clear that the subalgebra 
consisting of elements ${\bf T}_\alpha$ and ${\bf M}_{\beta\gamma}$ is 
isomorphic to $iso(3,1)$.
In addition, the subalgebra of $so(4,2)$ consisting of ${\bf K}_\alpha$ and 
${\bf M}_{\beta\gamma}$ is also isomorphic to the algebra $iso(3,1)$. 
The group of translations is an invariant subgroup of $ISO(3,1)$ and so the Lie 
group $ISO(3,1)$ is nonsemi-simple.   

The de Sitter and anti-de Sitter groups $SO(4,1)$ and $SO(3,2)$ are also 
subgroups of the conformal group $SO(4,2)$. These are simply the subgroups 
preserving the flat metrics with signatures $(\: + \: + \: + \: + \: - \: )$ 
and $(\: + \: + \: + \: - \: - \: )$ respectively. 
Thus, the algebra $so(4,2)$ contains the subalgebras $so(4,1)$ and $so(3,2)$, 
see appendix \ref{app:desitter}. 

Thus we have, for example, the following embedded subalgebras
\ba
so(4,2) \supset iso(3,1) \supset iso(3) \phantom{0} \supset so(3) \phantom{00} 
\phantom{,1} & & \nonumber\\ 
\supset so(3,1) \supset so(3) \phantom{00}  \phantom{,1} & & \nonumber\\  
so(4,2) \supset so(4,1) \, \supset so(4) \phantom{00} \supset 
so(3) \phantom{00}  \phantom{,1} & & \nonumber\\ 
\supset iso(3) \, \phantom{0} \supset 
so(3) \phantom{00}  \phantom{,1} 
& & \nonumber\\
\supset so(3,1) \supset so(3) \phantom{00}  \phantom{,1} & & \nonumber\\ 
so(4,2)  \supset so(3,2) \, \supset so(3,1) \supset so(3) 
\phantom{00}  \phantom{,1} & &
\ea
For more details of the subgroups of the conformal group $SO(4,2)$, see 
Beckers et al [10] and references therein.

\section{The de Sitter and anti-de Sitter Lie algebras}
\label{app:desitter}
Let us consider the Lie algebras of the isometry groups of the 
de Sitter and anti-de Sitter spacetimes.
For the Lie algebra $so(4,1)$ we can choose the following basis 
\ba
& & {\bf M}_{ij} = {\bf L}_{ij}; \nonumber\\
& & {\bf T}_i = {\bf L}_{i4} + {\bf L}_{i5}; \nonumber\\
& & {\bf K}_i = {\bf L}_{i4} - {\bf L}_{i5}; \nonumber\\
& & {\bf D} = {\bf L}_{45} \label{eqn:so41groupbasis}
\ea
where $i = 1,2,3$. We can write the Lie algebra in terms of the basis (\ref{eqn:so41groupbasis}) as  
\ba
& & [{\bf M}_{ij}, {\bf M}_{kl}] = 
\eta_{il}{\bf M}_{jk} + 
\eta_{jk}{\bf M}_{il} +
\eta_{ik}{\bf M}_{lj} +
\eta_{jl}{\bf M}_{ki}; \nonumber\\
& & [{\bf T}_i, {\bf T}_j] = 0; \qquad
[{\bf T}_i, {\bf M}_{jk}] = 
\eta_{ij}{\bf T}_k - \eta_{ik}{\bf T}_j;
\nonumber\\ 
& & [{\bf K}_i, {\bf K}_{j}] = 0; \qquad 
[{\bf K}_i, {\bf M}_{jk}]  =  
\eta_{ij}{\bf K}_k - \eta_{ik}{\bf K}_j;
\nonumber\\
& & [{\bf D}, {\bf K}_{i}] = {\bf K}_{i};\qquad
[{\bf D}, {\bf M}_{ij}] = 0; \qquad 
[{\bf T}_i, {\bf D}] = {\bf T}_i;  
\nonumber\\
& & [{\bf T}_i, {\bf K}_{j}] =  
2 (\eta_{ij}{\bf D} - {\bf M}_{ij}). 
\label{app:alg:so41}
\ea
In this basis it is clear that the elements $\{ {\bf T}_i, {\bf M}_{jk} \}$ 
form the Lie algebra $iso(3)$. 
We can also define
\ba
{\bf T}_i = {\bf L}_{i0} \nonumber\\
{\bf K}_i = {\bf L}_{i5} \nonumber\\
{\bf D} = {\bf L}_{05} 
\ea 
where $i = 1,2,3$ and $\alpha = 0 \equiv A=4$. We can then write the 
de Sitter and anti-de Sitter Lie algebras as
\ba
& & [{\bf T}_i, {\bf L}_{jk}] = \eta_{ij} {\bf T}_k - \eta_{ik} {\bf T}_j;
\qquad 
[{\bf T}_i, {\bf T}_j] = \mp {\bf L}_{ij}; \nonumber\\
& & [{\bf L}_{ij}, {\bf L}_{kl}] = 
\eta_{il}{\bf L}_{jk} + 
\eta_{jk}{\bf L}_{il} +
\eta_{ik}{\bf L}_{lj} +
\eta_{jl}{\bf L}_{ki}; \nonumber\\
& & [{\bf K}_i, {\bf L}_{jk}] = \eta_{ij} {\bf K}_k - \eta_{ik} {\bf K}_j;
\qquad [{\bf K}_i, {\bf K}_j] = {\bf L}_{ij}; \nonumber\\
& & [{\bf D}, {\bf L}_{ij}] = 0; \qquad 
[{\bf T}_i, {\bf D}] = \pm {\bf K}_i; \qquad
[{\bf K}_i, {\bf D}] = {\bf T}_i; \nonumber\\
& & [{\bf T}_i, {\bf K}_j] = - \eta_{ij} {\bf D};
\label{app:alg:so41:so32}
\ea
where we take the upper signs for $so(4,1)$ and the lower signs for $so(3,2)$. 
Thus we can see that the elements ${\bf T}_i$ and ${\bf L}_{jk}$ form the 
subalgebra $so(4)$ or $so(3,1)$ in the cases $so(4,1)$ and $so(3,2)$ 
respectively. In both cases the elements ${\bf K}_i$ and ${\bf L}_{jk}$ form 
the subalgebra $so(3,1)$.

\section*{References}
 
\noindent 1. Levine, J. {\it Groups of motions in conformally flat spaces.}
Bulletin of the American Mathematical Society, volume 42, 418-422 (1936). 

\noindent 2. Levine, J. {\it Groups of motions in conformally flat spaces. II.}
Bulletin of the American Mathematical Society, volume 45, 766-773 (1939). 

\noindent 3. Maartens, R. and Maharaj, S.D. {\it Conformal Killing vectors in Robertson-Walker spacetimes}, Classical and Quantum Gravity, vol 3,  
1005-1011 (1986).

\noindent 4. Maharaj, S.D. and Maartens, R. {\it General solution of 
Liouville's Equation in Robertson-Walker spacetimes}, General Relativity and 
Gravitation, vol 19, No. 12, 1217-1222 (1987).

\noindent 5. Maartens, R. and Maharaj, S.D. {\it Invariant solutions of 
Liouville's Equation in Robertson-Walker spacetimes}, General Relativity and 
Gravitation, vol 19, No. 12, 1223-1234 (1987).

\noindent 6. Hall, G.S. {\it Conformal symmetries in
general relativity.} Proceedings of the 3rd Hungarian Relativity Workshop, 
Tihany, (1989).

\noindent 7. Hawking, S.W. and Ellis, G.F.R. {\it The large scale structure 
of space-time.}, Cambridge University Press, (1973).

\noindent 8. Torrence, R.J. and Couch, W.E.   
{\it Note on Equivalence of Cosmological Space-times.} 
General Relativity and Gravitation, vol 18, No. 6, 585-589 (1986).

\noindent 9. Stephani, H. {\it General Relativity - an introduction to the 
theory of the gravitational field}, Cambridge University Press, second edition
(1990). 

\noindent 10. Beckers, J., Harnad, J., Perroud, M. and Winternitz, P.  
{\it Tensor fields invariant under subgroups of the conformal group of 
space-time.} J. Math. Phys., vol 19(10), 2126-2153 (1978).

\newpage
\begin{table*}[t]
\begin{center}{
\begin{tabular}{|l|l|r|}     \hline 
\emph{CKV} & \emph{conformal factor} & \emph{type \phantom{00}} \\ \hline
${\bf T}_\alpha$ & $\phi=0$ & KV \phantom{SC} \\
${\bf M}_{\alpha\beta}$ & $\phi=0$ & KV \phantom{SC} \\ 
${\bf D}$ & $\phi=1$ & HKV \phantom{S}\\
${\bf K}_\alpha$ & $\phi=2x'_\alpha$ & SCKV $\,$ \\ \hline
\end{tabular}
}\end{center}
\caption[]{\small{The conformal factors for the CKV of Minkowski 
spacetime.}}
\label{tab:minkowskickv}
\end{table*}
\begin{table*}[t]
 \begin{center}{
\begin{tabular}{|l|l|r|}     \hline 
\emph{CKV} & \emph{conformal factor} & \emph{type} \\ \hline
${\bf T}_0$ & $\phi= {1 \over 2} A S^{-1} {\partial S / \partial \tau} 
- {k \over 2} \epsilon K_- K_+^{-1}$ & CKV \\
${\bf T}_i$ & $\phi= -{k \over 2} K_+^{-1} {\bar x}^i  ( \epsilon S^{-1} 
{\partial S / \partial \tau} + (1-k\epsilon^2)^{1 \over 2})$ & CKV \\
${\bf M}_{0i}$ & $\phi= K_+^{-1} {\bar x}^i  ( -(1-k\epsilon^2)^{1 \over 2} 
S^{-1} {\partial S / \partial \tau} + k\epsilon)$ & CKV \\ 
${\bf M}_{ij}$ & $\phi=0$ & KV \\ 
${\bf D}$ & $\phi=K_-K_+^{-1} (\epsilon S^{-1} 
{\partial S / \partial \tau} + (1-k\epsilon^2)^{1 \over 2})$ & CKV \\
${\bf K}_0$ & $\phi= - 2S^{-1} {\partial S / \partial \tau} F^{-2} 
(2F \epsilon^2 K_- K_+^{-1}+(-\epsilon^2 +r^2) A)
-2 \epsilon K_- K_+^{-1}$ & CKV \\ 
${\bf K}_i$ & $\phi= 2S^{-1} {\partial S / \partial \tau} F^{-2} K_+^{-1} 
\epsilon {\bar x}^i 
(2F K_- K_+^{-1} + k (-\epsilon^2 +r^2)) + 2 {\bar x}^i K_+^{-1}$ & CKV \\ 
\hline
\end{tabular}
}\end{center}\caption[]{\small{The conformal factors for the CKV of 
Robertson-Walker spacetimes.}}
\label{tab:frwckv}
\end{table*}
\begin{table*}[t]
 \begin{center}{
\begin{tabular}{|l|l|r|}     \hline 
\emph{CKV} & \emph{conformal factor} & \emph{type} \\ \hline
${\bf P}_0$ & $\phi=S^{-1} {\partial S / \partial \tau}$ & CKV \\
${\bf H}$ & $\phi=K_- K_+^{-1} (S^{-1} {\partial S / \partial \tau} 
(1-k\epsilon^2)^{1 \over 2}-k \epsilon)$ & CKV \\ 
${\bf P}_i$ & $\phi=0$ & KV \\
${\bf {\bar Q}}^*_i$ & $\phi=-k K_+^{-1} {\bar x}^i (\epsilon S^{-1} 
{\partial S / \partial \tau} + (1-k\epsilon^2)^{1 \over 2})$ & CKV \\ \hline
\end{tabular}
}\end{center}\caption[]{\small{The conformal factors for certain linear 
combinations of Robertson-Walker CKV.}}
\label{tab:frwckv-newbasis}
\end{table*}
%
%
\begin{sidewaystable}
\centering
\setlength{\arrayrulewidth}{.8pt}
\begin{tabular}{|l|l|l|l|l|l|l|r|}     \hline 
\emph{} & \emph{${\bf H}\phantom{\biggl(}$} & \emph{${\bf P}_k$} & 
\emph{${\bf M}_{0k}$} 
& \emph{${\bf M}_{kl}$} & \emph{${\bf {\bar Q}}^*_k$} & \emph{${\bf P}_0$} 
& \emph{${\bf D}$} \\ \hline
${\bf H}\phantom{\biggl(}$ 
& 0 & $-k{\bf M}_{0k}$ & $-{\bf P}_k$ & 0 & 0 & $k{\bf D}$ & 
${\bf P}_0$\\ \hline
${\bf P}_i\phantom{\biggl(}$ & $k{\bf M}_{0i}$ & $-k{\bf M}_{ik}$ & $-\eta_{ik}{\bf H}$ &
 $\eta_{ik}{\bf P}_l - \eta_{il}{\bf P}_k$ & $-k\eta_{ik}{\bf D}$ & 0 & 
${\bf {\bar Q}}^*_i$\\ \hline  
${\bf M}_{0i}\phantom{\biggl(}$ & ${\bf P}_i$ & $\eta_{ik}{\bf H}$ & ${\bf M}_{ik}$ & 
$\eta_{ik}{\bf M}_{0l} - \eta_{il}{\bf M}_{0k}$ & $\eta_{ik}{\bf P}_0$
& ${\bf {\bar Q}}^*_i$ & 0 \\ \hline  
${\bf M}_{ij}\phantom{\biggl(}$ & 0 & $\eta_{jk}{\bf P}_i - \eta_{ik}{\bf P}_j$ 
& $\eta_{jk}{\bf M}_{0i} - \eta_{ik}{\bf M}_{0j}$ & $so(3)$ 
& $\eta_{jk}{\bf {\bar Q}}^*_i - \eta_{ik}{\bf {\bar Q}}^*_j$
& 0 & 0\\ \hline
${\bf {\bar Q}}^*_i\phantom{\biggl(}$ & 0 & $k\eta_{ik}{\bf D}$ & $-\eta_{ik}{\bf P}_0$ 
& $\eta_{ik}{\bf {\bar Q}}^*_l - \eta_{il}{\bf {\bar Q}}^*_k$ & $k{\bf M}_{ik}$ 
& $-k{\bf M}_{0i}$ & ${\bf P}_i$\\ \hline
${\bf P}_0\phantom{\biggl(}$ & $-k{\bf D}$ & 0 & $-{\bf {\bar Q}}^*_k$ & 0 & $k{\bf M}_{0k}$
& 0 & ${\bf H}$\\ \hline
${\bf D}\phantom{\biggl(}$ & $-{\bf P}_0$ & $-{\bf {\bar Q}}^*_k$ & 0 & 0 & $-{\bf P}_k$ 
& $-{\bf H}$ & 0\\ \hline
\end{tabular}
\caption[]{\small{The Lie algebra in terms of the basis of Maartens and 
Maharaj. $so(3)$ is shorthand for the commutation relations \\
$[{\bf M}_{ij}, {\bf M}_{mn}] =  
\eta_{jm}{\bf M}_{in} + 
\eta_{jn}{\bf M}_{mi} +
\eta_{im}{\bf M}_{nj} +
\eta_{in}{\bf M}_{jm}$.}}
\label{tab:frwckv-newbasis-algebra}
\end{sidewaystable}
%
%
\begin{sidewaystable}
\centering
\setlength{\arrayrulewidth}{.8pt}
\begin{tabular}{|l|l|l|l|l|l|l|r|} \hline 
\emph{} & \emph{${\bf P}_k\phantom{\biggl(}$} & \emph{${\bf M}_{kl}$} & 
\emph{${\bf M}_{0k}$} 
& \emph{${\bf K}_k$} & \emph{${\bf K}_0$} & \emph{${\bf P}_0$} 
& \emph{${\bf D}$} \\ \hline
${\bf P}_i \phantom{\biggl(}$ 
& $-k{\bf M}_{ik}$ 
& $\eta_{ik}{\bf P}_l - \eta_{il}{\bf P}_k$ 
& $-\eta_{ik}({\bf P}_0+{k \over 2}{\bf K}_0) $ 
& $2(\eta_{ik}{\bf D}-{\bf M}_{ik})$ 
& $2{\bf M}_{0i}$ 
& $0$ 
& ${\bf P}_i-{k \over 2}{\bf K}_i$
\\ \hline
${\bf M}_{ij}\phantom{\biggl(}$ 
& $\eta_{jk}{\bf P}_i - \eta_{ik}{\bf P}_j$ 
& $so(3)$ 
& $\eta_{jk}{\bf M}_{0i} - \eta_{ik}{\bf M}_{0j}$ 
& $\eta_{jk}{\bf K}_i - \eta_{ik}{\bf K}_j$ 
& $0$ 
& 0 
& $0$ \\ \hline
${\bf M}_{0i}\phantom{\biggl(}$ 
& $\eta_{ik} ({\bf P}_0 + {k \over 2} {\bf K}_0)$ 
& $\eta_{ik}{\bf M}_{0l} - \eta_{il}{\bf M}_{0k}$ 
& ${\bf M}_{ik}$ 
& $\eta_{ik}{\bf K}_0$ 
& ${\bf K}_i$
& ${\bf P}_i-{k \over 2} {\bf K}_i$ 
& 0 \\ \hline
${\bf K}_i \phantom{\biggl(}$ 
& $2({\bf M}_{ki}-\eta_{ik}{\bf D})$ 
& $\eta_{ik}{\bf K}_l - \eta_{il}{\bf K}_k$ 
& $-\eta_{ik}{\bf K}_0$ 
& 0 
& 0
& $2{\bf M}_{0i}$ 
& $-{\bf K}_i$ \\ \hline
${\bf K}_0 \phantom{\biggl(}$ 
& $-2{\bf M}_{0k}$ 
& 0 
& $-{\bf K}_k$ 
& 0
& 0  
& $2{\bf D}$ 
& $-{\bf K}_0$\\ \hline
${\bf P}_0 \phantom{\biggl(}$ 
& 0 
& 0 
& $-({\bf P}_k-{k \over 2} {\bf K}_k)$
& $-2{\bf M}_{0k}$ 
& $-2{\bf D}$
& 0
& ${\bf P}_0+{k \over 2} {\bf K}_0$
\\ \hline
${\bf D} \phantom{\biggl(}$ 
& $-({\bf P}_k-{k \over 2}{\bf K}_k)$ 
& 0 
& 0 
& ${\bf K}_k$
& ${\bf K}_0$ 
& $-({\bf P}_0+{k \over 2} {\bf K}_0)$ 
& 0\\ \hline
\end{tabular}
\caption[]{\small{The Lie algebra in terms of the continuous basis.}}
\label{tab:banana}
\end{sidewaystable}
%
%
\setlength{\arrayrulewidth}{.8pt}
\begin{sidewaystable}
\centering
\begin{tabular}{|l|l|l|l|l|l|l|r|} \hline 
\emph{} & \emph{${\bf H} \phantom{\biggl(}$} & \emph{${\bf P}_k$} & 
\emph{${\bf M}_{0k}$} & \emph{${\bf M}_{kl}$}
& \emph{${\bf K}_k$} & \emph{${\bf K}_0$}  
& \emph{${\bf D}$} \\ \hline
${\bf H} \phantom{\biggl(}$ 
& 0 
& $-R{\bf M}_{0k}$ 
& $-{\bf P}_k$ 
& 0 
& $-2{\bf M}_{0k}$ 
& $-2{\bf D}$ 
& ${\bf H}-{R \over 2}{\bf K}_0$
\\ \hline
${\bf P}_i \phantom{\biggl(}$ 
& $R{\bf M}_{0i}$
& $-R{\bf M}_{ik}$ 
& $-\eta_{ik}({\bf P}_0 + {R \over 2}{\bf K}_0)$ 
& $\eta_{ik}{\bf P}_l - \eta_{il}{\bf P}_k$ 
& $2(\eta_{ik} {\bf D} - {\bf M}_{ik})$ 
& $2{\bf M}_{0i}$ 
& ${\bf P}_i-{R \over 2} {\bf K}_i$ \\ \hline
${\bf M}_{0i}\phantom{\biggl(}$ 
& ${\bf P}_i$ 
& $\eta_{ik}({\bf P}_0 +{R \over 2}{\bf K}_0)$ 
& ${\bf M}_{ik}$ 
& $\eta_{ik}{\bf M}_{0l}-\eta_{il}{\bf M}_{0k}$ 
& $\eta_{ik}{\bf K}_0$
& ${\bf K}_i$ 
& 0 \\ \hline
${\bf M}_{ij}\phantom{\biggl(}$ 
& 0 
& $\eta_{jk}{\bf P}_i - \eta_{ik}{\bf P}_j$ 
& $\eta_{jk}{\bf M}_{0i} - \eta_{ik}{\bf M}_{0j}$ 
& $so(3)$ 
& $\eta_{jk}{\bf K}_i - \eta_{ik}{\bf K}_j$
& 0 
& 0 \\ \hline
${\bf K}_i \phantom{\biggl(}$
& $2{\bf M}_{0i}$ 
& $2({\bf M}_{ki}-\eta_{ik}{\bf D})$ 
& $-\eta_{ik}{\bf K}_0$ 
& $\eta_{ik}{\bf K}_l - \eta_{il}{\bf K}_k$
& 0  
& 0
& $-{\bf K}_i$\\ \hline
${\bf K}_0 \phantom{\biggl(}$ 
& $2{\bf D}$ 
& $-2{\bf M}_{0k}$ 
& $-{\bf K}_k$
& 0 
& 0
& 0
& $-{\bf K}_0$
\\ \hline
${\bf D} \phantom{\biggl(}$ 
& $-({\bf H}-{R \over 2}{\bf K}_0)$ 
& $-({\bf P}-{R \over 2}{\bf K}_k)$ 
& 0 
& 0
& ${\bf K}_k$ 
& ${\bf K}_0$ 
& 0\\ \hline
\end{tabular}
\caption[]{\small{The Lie algebra in terms of the de Sitter basis.}}
\label{tab:frwckv-desitter}
\end{sidewaystable}
%
\clearpage
\newpage
\clearpage
\newpage
\clearpage
\newpage

\end{document}